\documentstyle[prl,aps,epsf]{revtex}
\begin{document}
\draft
\twocolumn[
\hsize\textwidth\columnwidth\hsize\csname@twocolumnfalse\endcsname
\preprint{}
\title{Local Quasiparticle States near a Unitary Impurity with Induced
Magnetic Moment in a $d$-Wave Superconductor}
\author{Jian-Xin Zhu and C. S. Ting}
\address{
Texas Center for Superconductivity and Department of Physics,
University of Houston, Houston, Texas 77204
}
\maketitle
\begin{abstract}
{ 
The local quasiparticle density of states around a unitary impurity with 
a Kondo-like  magnetic moment induced at its nearest neighbors in a
$d$-wave superconductor is studied  within  the slave-boson mean-field
approach. The Hamiltonian is exactly diagonalized and the results are
obtained self-consistently.  We show that the interference between the
strong  impurity potential scattering and the Kondo effect leads to  novel
quasiparticle spectra  around the impurity, which  are strikingly
different from the case of a single unitary or magnetic impurity.  The
recent STM image of the local differential  tunneling  conductance around
a Zn impurity in a high-$T_c$ cuprate can  be essentially explained  if
the blocking effect of BiO surface layer between the tip and probed CuO$_2$
plane [Zhu {\em et al.}, Phys. Rev. B {\bf 62}, 6207 (2000)] is taken
into account.
}
\end{abstract} 
]

\narrowtext

The scanning tunneling microscopy (STM) turns out to be a powerful probe 
of the local effects of individual impurity atoms or defects on the
fundamental properties of high-$T_c$ cuprates~\cite{Pan00,Huds99,Yazd99}. 
The experiments revealed the induction of a virtual bound quasiparticle
state around the individual impurity, providing
a strong evidence of the $d$-wave pairing symmetry of high-$T_c$
superconductors~\cite{Harl95}. This observation is motivated by
the earlier predictions of Balatsky, Salkola and 
co-workers~\cite{Bala95,Salk96}. 
However, the high resolution STM imaging of the effect of individual 
impurity atoms Zn substituted with a controlled manner for Cu in
Bi$_2$Sr$_2$CaCu$_2$O$_{8+\delta}$ (BSCCO)~\cite{Pan00} also presented an
unexpected 
spatial distribution of the local density of states (LDOS) strength at the
resonance energy: It has a strongest intensity 
directly on or above the Zn site, local minimum
on its nearest-neighboring sites and local maximum on
second-nearest-neighboring
sites. This distribution is opposite to all theories based on the 
strong atomic-like nonmagnetic impurity  
model~\cite{Bala95,Salk96,Tsuc99,Zhu00a}, which show a vanishing density
of states directly at the impurity site  and a strong resonance peak  on its
nearest neighbors. Recent NMR measurements of Zn and Li
doped  YBa$_2$Cu$_3$O$_{6+y}$~\cite{Allo91,Juli00,Bobr00} and the
specific heat measurement of Zn-doped 
YBa$_2$Cu$_3$O$_{6.95}$~\cite{Siss00}  indicated that a Kondo like
magnetic moment with $S=\frac{1}{2}$ is induced at the nearest neighboring
sites of the Zn (or Li) impurity. Motivated by the latter  experiments,
the Kondo effect on the local quasiparticle states around an induced
magnetic moment in a $d$-wave
superconductor has been theoretically studied by several
 groups~\cite{Polk00,Zhu00c,Zhang00}. 
All these results show wide-distanced double-peak structure with respect
to the zero bias at the impurity sites, and therefore
do not agree with the STM  imaging of the effects of a Zn impurity.  
We believe that these studies seem to be more suitable to
describe the Kondo impurities such as Mn and Co substituted for Cu 
in high-$T_c$ cuprates instead of Zn or even Ni whose spin is believed to
have a ferromagnetic coupling with conduction 
electrons~\cite{Khal97,Salk97,Tsuc00}.  In order to compare with experiments,
 the scattering of quasiparticles due to both Zn and the induced magnetic 
moment around it needs to be carefully investigated.  
So far this problem has not been properly addressed, particularly in view
of the 
scattering strength of a Zn impurity being in the unitary limit as it will
be estimated below. Valence counting suggests that  the Zn impurity maintains
a nominal Cu$^{2+}$ charge, 
the Zn$^{2+}$ [$3d^{10}$] with $S=0$ configuration indicates it would act as
a nonmagnetic impurity. According to  Ref.~\cite{Hsu},  the second
and third 
ionization energies for the Zn element are respectively 18~eV and 39.8~eV, 
while  the third ionization energy of the Cu element is 36.9~eV. In the 
cuprates, the relevant  $3d$ Cu$^{2+}$  electrons form a narrow
$d_{x^2-y^2}$ band with
bandwidth less than 2~eV. The center
of the band, where the chemical potential lies, is located around -36.9~eV 
below the vacuum. The top of 
$3d^{10}$ level of a Zn$^{2+}$ ion is at $-39.8~\mbox{eV}$ which is below
the bottom of the $d_{x^2-y^2}$ band.
The conduction electrons near the chemical potential will experience a 
repulsive  local potential $V_0\approx (-18~\mbox{eV}) - (-36.9~\mbox{eV}) 
=18.9~\mbox{eV}$ at the Zn site.  For such a large local potential
as compared to the bandwidth, Zn  can indeed be regarded as a unitary
impurity. Other evidence for the strong potential scattering from Zn 
has also been provided experimentally~\cite{Chien91}. 
It is important to point out that our present consideration  
is very different from the case studied by the work~\cite{Polk00} 
where the authors assumed the Zn impurity to have a  
weak attractive potential.

In this Letter we study the local quasiparticle spectrum around a 
unitary impurity with a Kondo-like magnetic moment induced 
at its nearest neighbors 
in a $d$-wave superconductor.  Based on this more sophisticated model, it is 
shown that the presence of the strong potential scattering from the 
nonmagnetic impurity leads to a stronger single resonant peak in the LDOS
directly at the site of  the magnetic moment,
in contrast to the case for a single magnetic 
impurity~\cite{Zhu00c,Zhang00}. Comparing with
the case of a single nonmagnetic unitary impurity, the LDOS is strongly 
enhanced by the Kondo effect from the induced magnetic moment. We also
show that  the novel pattern observed in the 
STM experiments can be qualitatively explained by taking into account 
the blocking effect~\cite{Zhu00b} 
of the BiO surface layer which exists between the tunneling tip and the
CuO$_2$ plane where the Zn impurities are located.
   
We model the $d$-wave superconductor defined on a two-dimensional 
lattice within a phenomenological BCS framework. The unitary impurity is
taken  to be  at the origin ${\bf r}_{0}=(0,0)$. 
Nearest-neighboring to the impurity, the induced magnetic moment  
is described by the Anderson $s$-$d$ exchange model. 
Since we are most interested in 
the low temperature quasiparticle physics, the on-site repulsive 
interaction on the moment-sitting site is assumed to be infinite 
so that the double occupancy on this 
site is forbidden.  We use the slave-boson 
mean-field approach~\cite{Barn76,Read83} 
to study  the Kondo effect associated with the induced
magnetic moment.
The system Hamiltonian can then be diagonalized
by solving the the Bogoliubov-de Gennes equations~\cite{Zhu00c}:
\begin{equation}
\sum_{\bf j}\left( \begin{array}{cc} 
H_{\bf ij} & \Delta_{\bf ij} \\
\Delta_{\bf ij}^{*} & -H_{\bf ij} 
\end{array} \right) 
\left( \begin{array}{c} 
u_{\bf j}^{n} \\ v_{\bf j}^{n} 
\end{array} \right)
=E_{n}
\left( \begin{array}{c} 
u_{\bf i}^{n} \\ v_{\bf i}^{n} 
\end{array} \right)\;,
\label{EQ:BdG}
\end{equation}
with the bosonic number $b_0$ and  
the $d$-wave bond pairing amplitude $\Delta_{\bf ij}$ 
subject to the self-consistent condition 
\begin{equation}
\Delta_{\bf ij}=\frac{g_{\bf ij}}{2}\sum_{n}(u_{\bf i}^{n}v_{\bf j}^{n*}
+u_{\bf j}^{n} v_{\bf i}^{n*})\tanh(E_{n}/2k_{B}T)\;,
\end{equation}
and
\begin{equation}
b_{0}^{2}=1-2\sum_{n}\{ \vert u_{{\bf r}_{1}}^{n}\vert^{2} f(E_{n}) 
+\vert v_{{\bf r}_{1}}^{n}\vert^{2}[1-f(E_{n})]\}\;.
\label{EQ:SINGLE}
\end{equation}
Here $\left( \begin{array}{c} u_{\bf i}^{n} \\v_{\bf i}^{n} 
\end{array} \right)$
is the 
quasiparticle wavefunction with eigenvalue $E_{n}$.
$H_{\bf ij}=-\tilde{t}_{\bf ij} -\mu \delta_{{\bf i}\neq {\bf r}_{1},{\bf
j}}+V_{0}\delta_{{\bf i}{\bf r}_{0}}+
(\epsilon_{d}+\lambda_0)\delta_{{\bf i}{\bf r}_{1}}\delta_{\bf ij}$
is the single particle Hamiltonian. 
Without loss of generality, we temporarily assume the magnetic moment to be at 
${\bf r}_{1}=(1,0)$, $\boldmath{\mbox{$\delta$}}$ are the unit vectors
of the square lattice, $\tilde{t}_{\bf ij}
={\cal V}b_{0}$ for $({\bf ij})
=({\bf r}_{1}+\boldmath{\mbox{$\delta$}}, {\bf r}_{1})$ 
or $({\bf r}_{1},{\bf r}_{1}+\boldmath{\mbox{$\delta$}})$ 
and $t$ otherwise with ${\cal V}$ being the coupling strength between
the moment and conduction electrons and $t$ the hopping integral
between conduction electrons,  
$\mu$ is the chemical potential, 
$V_{0}$ is the strength of impurity potential, $\epsilon_{d}$ is the 
bare energy level of the moment electron with respect to the chemical 
potential for the band electrons for superconductivity, 
$\lambda_0$ is the Lagrange multiplier introduced to enforce the 
single occupancy constraint. 
Here $g_{\bf ij}$ represents the strength of $d$-wave pairing 
interaction, the value of which is 
$g_{\bf ij}
=0$ for $({\bf ij})=({\bf r}_{1}+\boldmath{\mbox{$\delta$}}, {\bf r}_{1})$ 
or $({\bf r}_{1},{\bf r}_{1}+\boldmath{\mbox{$\delta$}})$ upon the assumption
of different nature of the conduction and moment electrons~\cite{Note1} and 
constant $g$ otherwise. Finally, $f(E)=[1+\exp(E/k_{B}T)]^{-1}$ is the Fermi 
distribution function.

Throughout the work,   
our numerical calculation is performed on a square lattice 
with size $N_{L}=N_x\times N_y=35\times 35$ and averaged over 36 
wavevectors in the supercell Brillouin zone. 
The strong scattering potential of the unitary impurity is taken to
be $V_{0}=100$, the bare energy level $\epsilon_d=-2$,
the chemical potential $\mu=-0.2$, and the pairing interaction 
$g=1.2$. Unless specified otherwise, hereafter we measure the energy 
in units of the hopping intergal $t$ and
the length of the lattice constant $a$. 
A trivial solution to Eq.~(\ref{EQ:BdG}) with 
$\lambda_0=-\epsilon_{d}$
and $b_0=0$ is always found, which describes the moment decoupled
from the conduction electrons. However, the physically desired value 
of these two parameters are determined by minimizing the free 
energy with the procedure as detailed in Ref.~\cite{Zhu00c}.
With the chosen parameter values, we find a critical value 
of the coupling strength ${\cal V}_{c}\approx 0.9$, below which the moment 
is decoupled from the conduction band.  Compared with the 
single Kondo impurity case, the larger ${\cal V}_{c}$ obtained 
here is due to the fact that the conduction electron density 
is strongly depressed on the unitary impurity site 
nearest-neighboring to the moment so that the screening effect 
is weakened. When ${\cal V}>{\cal V}_{c}$, this free local
spin state gives way to  the Kondo screened state.
For ${\cal V}=1.0$, it is found that $\lambda_0=2.15$ and 
$b_0=0.56$. The self-consistent solution also gives a $d$-wave 
order parameter which is strongly suppressed around the unitary 
impurity in the scale of the coherence $\xi_0\equiv \hbar v_{F}/\pi 
\Delta_g$, where $v_{F}$ is the Fermi velocity and 
$\Delta_g\approx 0.4$ is the bulk $d$-wave energy gap.     

Once the self-consistent solution is obtained, the LDOS is 
then calculated according to 
\begin{equation}
\rho_{\bf i}=2\sum_{n}[\vert u_{\bf i}^{n}\vert^{2} \delta(E_{n}-E) 
+\vert v_{i}^{n} \vert^{2} \delta(E_{n}+E)]\;,
\end{equation}
which is proportional to the local differential tunneling conductance 
as measured by the STM experiments. Our numerical result indicates  that 
the LDOS is vanishing at the unitary impurity site, 
consistent with previous calculations by others~\cite{Salk96,Tsuc99}.
Figure~\ref{FIG:LDOS1} shows the LDOS as a function of energy 
on sites nearest-neighboring to the unitary impurity site, $A=(1,0)$ 
[i.e., the magnetic moment-sitting site], $B=(0,1)$, $C=(-1,0)$, $D=(0,-1)$
and on the corner site of the supercell.
Due to the spatial inverse symmetry with respect to  the co-linear 
axis connecting the impurity and moment sites, the LDOS on sites 
$B$ and $D$ are identical. Notice that the LDOS on the corner site of
the supercell recovers the bulk density of states of a clean $d$-wave 
superconductor, which indicates that the supercell size and the 
number of wave vectors in the Brillouin zone has been 
large enough to uncover the physics intrinsic to an isolated 
impurity with an induced moment. The LDOS on the nearest-neighboring 
sites has evidenced the existence of near-zero energy resonant states 
by showing the near-zero energy peaks. Compared with a single nonmagnetic
unitary impurity or a single Kondo impurity, the unitary impurity with 
an induced moment shows much richer characteristics of the quasiparticle
resonant states in its vicinity: (i) In strikingly contrast to a single Kondo 
impurity case~\cite{Zhu00c,Zhang00}, the LDOS on the magnetic moment-sitting 
site-$A$ only displays a strong single near-zero-energy peak, which comes 
in a unique manner from the presence of the unitary impurity. (ii) 
Different from a single nonmagnetic unitary impurity 
case~\cite{Salk96,Zhu00a}, the LDOS 
on site-$B$ and site-$D$ exhibits the splitting of the 
near-zero-energy peak. This difference is mainly due to the presence of 
the moment near the unitary impurity, which plays the role 
of weak scattering center and reduces the effective hopping 
integral locally. As a comparison, we have also 
considered a (10)-oriented dimer consisting of
two nearest-neighboring unitary impurities
and found that 
the splitting is much increased. Therefore, in the Kondo resonance
regime, the unitary impurity plus a nearby moment composite 
can be effectively regarded a tailed nonmagnetic unitary impurity.  
Since the influence of the induced moment is very weak on site-$C$, 
the LDOS on this site only shows a single near-zero energy peak. 
(iii) As a result of the Kondo resonance, the intensity of the 
near-zero-energy peak on site-$A$ is about four times large as that on 
other three nearest-neighboring sites of the unitary impurity.  
So far we have considered the local quasiparticle resonant states 
around the unitary impurity with a moment induced statically. 
When a Zn impurity is substituted for Cu in a high-$T_c$ cuprate, 
the $S=\frac{1}{2}$  moment associated with Cu$^{2+}$ is removed from the
system. As a
result the local 
antiferromagnetic pairing is broken and a moment with spin-$\frac{1}{2}$ 
is induced at any of  the four
neighboring sites of the impurity with equal probability.  
Our treatment here is consistent with the NMR 
experimental 
indication that {\em the Zn impurity only reveals
already-existing  moments, localized on Cu sites of the doped
antiferromagnet}~\cite{Juli00} {\em and the induced spins on the nearest
neighboring copper sites are not correlated}~\cite{Ishida97}.
It is different from the approach based on a   
putatively single moment around the Zn impurity like 
a spin cluster.
After  averaging over the four moment-position configurations around 
the impurity, the combined
contribution from  sites-$A$ through $D$ leads to the averaged 
LDOS $\rho_{\bf i}^{\mbox{\tiny avg}} = (\rho_{A} + \rho_{B}+ 
\rho_{C} + \rho_{D} )/4$ at  any of these four  sites.
As shown in Fig.~\ref{FIG:LDOS2}, $\rho_{\bf i}^{\mbox{\tiny avg}}$  
is characterized by a very strong sharp peak close  to the zero 
energy.  Compared to the case of a single nonmagnetic unitary
impurity, the corresponding peak intensity is enhanced by about 40\%. 
Also interestingly, the LDOS spectrum also shows a narrow side-peak 
with a rather weak intensity. The highly asymmetric double-peak 
structure  distinguishes it with a sharp and narrow dip in between and their
separation  is only about 5\% of $2\Delta_g$. In comparison, the double
peaks obtained for a single Kondo impurity 
has a separation as large as 50\% of $2\Delta_g$~\cite{Zhu00c}.  

For the purpose to understand  the LDOS spectrum and the spatial
distribution 
of its strength at the resonance bias observed in
the STM experiment, we need to know the  nature of the surface layer in
BSCCO on which the tip is
directly probing. There are ample experimental evidences that the cleaved 
surface along the $c$-axis is 
always the BiO layer. The superconducting CuO$_2$ layer on which  Zn
impurities are located and the 
STM tip would like to explore, lies about $5\AA$ below the surface BiO
layer. The crystal structure shows that each Bi atom sits 
vertically on the top of each Cu atom or Zn atom, and there is 
also an O atom associated with SrO layer between them. 
When the STM tip is trying to probe the Zn site, 
the hard core of the  Bi$^{3+}$ ion (with radius $1.03~\AA$) and also
O$^{2-}$ (with radius $1.4~\AA$)
sitting on top of it will inevitably
block~\cite{Zhu00b} the tunneling current from 
directly passing through the Zn site, where Zn$^{2+}$ has the ionic radius
$0.74~\AA$.
Because the STM image has such a high spatial resolution to the atomic scale, 
the tunneling current from the tip  to the CuO$_2$ plane is only 
distributed within a small area, 
the linear  dimension of which is about one lattice constant. Therefore, 
the measured LDOS 
above  the {\em probed} Zn or Cu site is mainly contributed from its four    
nearest-neighboring sites, which can be approximately expressed as 
$\rho_{\bf i}^{\mbox{\tiny exp}}
=A\sum_{\boldmath{\mbox{$\delta$}} } \rho^{\mbox{\tiny avg}}_{{\bf
i}+\boldmath{\mbox{$\delta$}} } $. Here $A<1$ is a constant depending on
the distance between the tip and the four nearest-neighboring 
Cu atoms.  
As a result, the STM spectrum on
the Zn site observed by experiment is in good agreement with  
$\rho_{\bf i}^{\mbox{\tiny exp}}$  
at ${\bf i}=(0,0)$ which has the identical  characteristic as 
that shown in Fig.~\ref{FIG:LDOS2}. 
Experimentally, a small-side peak adjacent  to the left of the strong 
resonant peak was 
indeed observed in the STM measurement~\cite{Pan00}. 
This is consistent with our result in Fig. 2 and it 
could very well be a signature of the induced Kondo-like magnetic
moment around the impurity. In Fig.~\ref{FIG:BUBBLE}, a bubble plot is given
 for the spatial 
distribution of $\rho_{\bf i}^{\mbox{\tiny exp}}$ at the near-zero-bias
resonance energy $E=0.02$, where the size of each black dot is
proportional to the LDOS intensity at that lattice
site. As it is shown, the $\rho_{\bf i}^{\mbox{\tiny 
exp}}$ distribution agrees, at least qualitatively with the
measured differential-tunneling conductance image~\cite{Pan00}. 
  
To conclude, we have considered the effect of an individual
nonmagnetic unitary impurity with induced magnetic moment 
in a $d$-wave superconductor. To be applicable to the Zn impurity
substituted for Cu in copper-oxide cuprates, the model has taken care of 
both the strong potential scattering from the nonmagnetic impurity 
itself and the Kondo effect associated with the induced moment. 
We find that the induced magnetic moment strongly enhances the local
quasiparticle density of states near the impurity. 
In addition, the blocking effect of the BiO surface layer 
between the STM tip and the probed CuO plane is essential to explain the
observed resonance peak on the Zn site and 
the spatial pattern of the differential tunneling conductance around a Zn
impurity in BSCCO. 
   
{\bf Acknowledgments}: We wish to thank  C. R. Hu for useful discussions. 
 This work was supported by the Texas Center for 
Superconductivity at the University of Houston though the State of Texas, 
by a grant from the Robert A. Welch Foundation, 
by a Texas ARP grant (003652-0241-1999), 
and by NSF-INT-9724809.

\begin{figure}
\caption{The local density of states on the nearest-neighboring sites
of the unitary impurity $A=(1,0)$ (a), $B=(0,1)$
(b), and $C=(-1,0)$ (c). Also shown with the dashed line the LDOS on the
corner site of the supercell, which resembles the bulk DOS for a clean
$d$-wave superconductor. Here the $A$ site is the position of the induced
moment with the nonmagnetic unitary impurity located at (0,0)-site.
}
\label{FIG:LDOS1}
\end{figure}

\begin{figure}
\caption{The local density of states on the nearest-neighboring
site of the nonmagnetic unitary impurity after the average over the 
four moment-position configurations. As a comparison, also shown with
the dashed line the LDOS on the corresponding site  
for a single  nonmagnetic unitary impurity (i.e., without induced
magnetic moment). 
}
\label{FIG:LDOS2}
\end{figure}

\begin{figure}
\caption{A two-dimensional bubble view of the spatial distribution of the
local density of states at the resonance energy $E=0.02$ around the
nonmagnetic unitary impurity with a magnetic moment. The bubble size on
each Cu site scales the LDOS intensity. 
}
\label{FIG:BUBBLE}
\end{figure}

\end{document}